\documentclass[fleqn,usenatbib]{mnras}

\usepackage{newtxtext,newtxmath}
\usepackage{xcolor}
\usepackage{multirow}

\usepackage[T1]{fontenc}

\DeclareRobustCommand{\VAN}[3]{#2}
\let\VANthebibliography\thebibliography
\def\thebibliography{\DeclareRobustCommand{\VAN}[3]{##3}\VANthebibliography}

\usepackage{graphicx}	
\usepackage{amsmath}	

\title[VHE gamma rays from GRBs with EC emission]{Very high-energy gamma rays from GRB~180720B and GRB~190829A with external Compton emission}

\author[M. Barnard, S. Razzaque and J. C. Joshi]{
Monica Barnard,$^{1}$\thanks{E-mail: monicabarnard77@gmail.com, monicab@uj.ac.za}
Soebur Razzaque,$^{1,2,3}$\thanks{E-mail: srazzaque@uj.ac.za}
and Jagdish C. Joshi $^{4,1}$\thanks{E-mail: jagdish@aries.res.in}
\\
$^{1}$Centre for Astro-Particle Physics (CAPP), Department of Physics, University of Johannesburg, PO Box 524, Auckland Park 2006, South Africa\\
$^{2}$Department of Physics, The George Washington University, Washington, DC 20052, USA \\
$^{3}$ National Institute for Theoretical and Computational Sciences (NITheCS), South Africa \\ 
$^{4}$ Aryabhatta Research Institute of Observational Sciences (ARIES), Manora Peak, Nainital 263001, India\\
}

\date{Accepted XXX. Received YYY; in original form ZZZ}

\pubyear{2023}

\begin{document}
\label{firstpage}
\pagerange{\pageref{firstpage}--\pageref{lastpage}}
\maketitle

\begin{abstract}
Gamma-ray bursts (GRBs) comprise of short, bright, energetic flashes of emission from extragalactic sources followed by a longer afterglow phase of decreased brightness. Recent discoveries of very-high-energy (VHE, $\gtrsim 100$~GeV) afterglow emission from GRB 180720B and GRB 190829A by H.E.S.S.\ have raised questions regarding the emission mechanism responsible. We interpret these observed late-time emission to be the result of inverse Compton emission of ultra-relativistic electrons in the GRB blastwave in an external radiation field, i.e., external Compton (EC), considering both the wind and interstellar medium scenarios. We present predictions of multiwavelength light curves and energy spectra, ranging from optical to VHE, and include the synchrotron and synchrotron self-Compton (SSC) radiation mechanisms as well. We corrected the EC and SSC model for the $\gamma$-ray attenuation by absorption of photons through their interaction with the extragalactic background light (EBL). We compared our results to multiwavelength data and found that EC gives a satisfactory fit for a given set of fixed model parameters for GRB~180720B, whereas SSC result in a better fit for GRB~190829A. For both GRBs a wind environment is preferred over constant density inter-stellar medium, and the Cosmic Microwave Background as the external radiation field. However, with more data and an effective optimisation tool we can find a more robust fit of the model, implying better constraints on the GRB environment and the particle energy requirements for the emission observed at late times. This has consequences for future observations of GRBs at these extreme energies.
\end{abstract}

\begin{keywords}
(transients:) gamma-ray bursts -- (stars:) gamma-ray burst: individual: 180720B -- (stars:) gamma-ray burst: individual: 190829A
\end{keywords}



\section{Introduction}


Gamma-ray burst (GRB) afterglow radiation is characterised by their diverse emission across broad wavelengths, making them fascinating multiwavelength emitters \citep[see, for reviews,][]{2004RvMP...76.1143P, 2006RPPh...69.2259M, 2015PhR...561....1K}. Their multiwavelength features can be explained using the external-shock models by including detail processes, i.e., reverse shock, structured jets, arbitrary circumburst medium density profile, etc \citep{2015ApJS..219....9W}. Synchrotron radiation in the external shocks is very useful to explain the radiation up to $\sim10$ GeV \citep[see, e.g.,][]{2011MNRAS.412..522B, 2013FrPhy...8..661G, 2022Galax..10...74G}. However, processes such as the inverse-Compton scattering \citep[see, e.g.,][]{1999ApJ699C, 2001ApJ548787S, 2001ApJ...559..110Z, 2008MNRAS1483F}, proton-synchrotron radiation \citep[see, e.g.,][]{2010OAJ.....3..150R, 2020A&A...636A..82G, 2022arXiv221002363I, 2023ApJ947L14Z} and photopion interactions \citep[see, e.g.,][]{1998ApJ...499L.131B, 2009ApJ...699..953A, 2020ApJ...895L..41S} can extend the spectrum of gamma-rays beyond 10 GeV, as expected in theoretical models. However, the confirmation of VHE signal in afterglows came recently by the ground-based Cherenkov telescopes; i.e., Major Atmospheric Gamma Imaging Cherenkov Telescopes (MAGIC) \citep{2019Natur575459M} and High Energy Stereoscopic System (H.E.S.S.) \citep{ Abdalla2021_GRB190829a}, respectively. The VHE signal from GRB 180720B \citep{Abdalla2019_GRB180720b}, GRB 190829A \citep{Abdalla2021_GRB190829a}, and GRB 190114C \citep{2019Natur575459M, 2019Natur575455M} have provided us with an opportunity to investigate the physical properties of these GRBs. 

The VHE signal has been modelled using synchrotron-self-Compton (SSC) mechanism for GRB 190114C by \citet{2019Natur575459M, 2019ApJ...884..117W, Joshi2021, 2021ApJ923135D, yama2022MNRAS142Y, 2023MNRAS.520..839K}. However, hadronic $p\gamma$ interaction models are also used to study this burst \citep{2020ApJ...895L..41S}. The VHE observations are also used to constrain the intergalactic magnetic field values \citep{2020PhRvD.101h3004W, 2023PhRvD.107f3030D} and the energetics and radiative efficiency of GRB fireball \citep{2023arXiv230206116L}.
SSC and $p\gamma$ processes have also been used to study the broadband radiation from GRB 180720B \citep{Fraija2019, 2020ApJ...895L..41S} and GRB 190829A \citep{Abdalla2021_GRB190829a, 2021ApJ91812F, 2021ApJ...917...95Z, 2022ApJ92970S}. Further, \citet{2022ApJ...925..182H} used the VHE afterglow emission to study the maximum energy of electrons accelerated in those blastwaves.
%
Recently, an unusually bright burst GRB 221009A, also termed as the brightest of all time (B.O.A.T.), has been discovered \citep{Dichiara_2022, Galanti2022, Huang_2022, Veres_2022} and its  afterglow has been detected from radio to VHE energies~\citep{2023ApJ24W, LHAASO2023Sci, 2023arXiv230314172L}. This burst was also followed by the IceCube Neutrino Observatory but not detected \citep{2023ApJ94626A}. Further, based on the multiwavelength observations, detection of $\sim 18$ TeV photons \citep{Huang_2022} has been used to explore new physics \citep{baktash22, 2023ApJ...942L..21F, 2023PhLB..83937824N, 2023APh...14802831L, Cheung2022arXiv221014178C, Smirnov2022arXiv221100634S}, GRB jet composition studies \citep{2023AnA...670L..12D, Batista2022arXiv221012855A, 2023ApJ943L2L, 2023ApJ...946L..23L, 2023ApJ947L14Z}, and jet structure \citep{2023MNRAS.522L..56S, 2023ApJ...947...53R} and jitter radiation as a possible mechanism for its production \citep{2023ApJ946.89M} etc.  

In this paper, we will focus on the origin of the VHE signal from GRB 180720B \citep{Abdalla2019_GRB180720b} and GRB 190829A \citep{Abdalla2021_GRB190829a} and study the role played by the external-Compton (EC) radiation mechanism and fit multiwavelength data. However, we will also include the contributions due to the synchrotron and SSC mechanism. In the EC scenarios, relativistic electrons in the blastwave upscatter photons from an external radiation field. Previously, the EC model has been applied for the VHE radiation from GRB 190829A by \citet{2021ApJ92055Z}, where they used \emph{Swift}-XRT detected flare photons at $\sim 10^3$ s, as target photons.
We considered the low energy seed photons to be from Cosmic Microwave Background (CMB) origin. In Section~\ref{sec:MWmodel} we briefly describe the blastwave evolution, particle transport, the different radiation mechanisms considered and the implementation thereof, as well as the absorption of sub-TeV photons in the blastwave. In Section~\ref{sec:results} we apply our model to the before mentioned GRBs and compare our results to the multi-wavelength observations, with the discussion following in Section~\ref{sec:discussion}.

\section{Multi-wavelength emission model} \label{sec:MWmodel}

\subsection{GRB blastwave evolution} \label{sec:blastwave} 

The GRB blastwave with isotropic-equivalent initial kinetic energy of $E_k$ (erg) and initial bulk Lorentz factor $\Gamma$ interacts with circumburst medium of density $n(R)$ and generates external shocks \citep{blandford_mckee_soln_76}. Here $R$ is the distance from the central engine. The blastwave 
accumulates matter from the surrounding environment and decelerates. Here we have only used the adiabatic expansion scenario in a constant-density interstellar medium (ISM) and in a wind environment. We have also used formulas and their numerical values from \citet{Joshi2021}. The blastwave deceleration time is then  
%
\begin{eqnarray}
t_{\rm dec} =
\left[\frac{17E_{k}(1+z)^3}{64\pi n m_p c^5 \Gamma^8}\right]^{1/3} \,;~~~~ {\rm ISM}
\label{eq:dec_time_ism}
\end{eqnarray}
in ISM, where $n(R) =n_0 ~{\rm cm^{-3}}$, and
\begin{eqnarray}
t_{\rm dec} = 
\frac{9E_{k}(1+z)}{16\pi A m_p c^3 \Gamma^4}  \,;~~~~ {\rm Wind}
\label{eq:dec_time_wind}
\end{eqnarray}
in wind, where $n(R) =A R^{-2} ~{\rm cm^{-3}}$. 
For a stellar wind of velocity $v_w = 10^8 v_8 ~{\rm cm~s^{-1}}$ with mass loss rate $\dot{M}_w = 10^{-5} \dot{M}_{-5} M_{\odot}~{\rm yr^{-1}}$, we have $A = \dot{M}_w/ (4 \pi v_w m_p) = 3.02 \times 10^{35} A_{\star}~{\rm cm^{-1}}$, and $A_{\star} \equiv \dot{M}_{-5}/v_8$.
The time evolution of the Lorentz factor of the shocked gas in the blastwave can be written in the ISM and wind scenario as \citep[see, e.g.,][]{2013PhRvD..88j3003R}
\begin{eqnarray}
\Gamma_{\rm g}(t) = \frac{\Gamma}{\zeta}
\begin{cases}  
 \left(\frac{t_{\rm dec}}{t}\right)^{3/8} \,; & {\rm ISM} \\
\left(\frac{t_{\rm dec}}{t}\right)^{1/4} \,; & {\rm Wind}
\end{cases}
\end{eqnarray}
where the parameter $\zeta$ depends on the details of the blastwave evolution, in particular on the similarity parameter \citep{blandford_mckee_soln_76}. For our calculations, we have used $\zeta = \sqrt{2}$. The corresponding radius of the blastwave evolves as 
\begin{eqnarray}
R(t) = 
  \begin{cases} 
  \frac{16\Gamma^{2}_{\rm g}(t)ct}{(1+z)} \,; & {\rm ISM} \\
  \frac{8\Gamma^{2}_{\rm g}(t)ct}{(1+z)} \,; & {\rm Wind}
 \end{cases}
\end{eqnarray}

\subsection{Particle acceleration and spectrum} \label{sec:transport}
A population of non-thermal electrons in the shocked gas is assumed to be accelerated in the magnetic field in the shocked region. The energy density of this magnetic field $u^\prime_B = B^{\prime 2}/8\pi$ is assumed to be a fraction $\epsilon_B$ of the energy density of the shocked gas in the blastwave frame, $u^\prime_{\rm g}=4n(R)m_pc^2 \Gamma_{\rm g}^2$, for a strong shock. Thus the magnetic field is
\begin{eqnarray} \label{eq:Bprime}
B^\prime = \sqrt{32 \pi \epsilon_{B} n(R) m_p c^2} \Gamma_{\rm g}.
\end{eqnarray}
The electrons also cool due to synchrotron and SSC radiation. The resulting electron Lorentz factor spectrum in the slow-cooling regime, isotropic in the blastwave frame, relevant for late afterglow emission, is given by
\begin{eqnarray}
n^\prime(\gamma^\prime) = \frac{k}{4\pi} \times
\begin{cases}
  (\gamma^\prime/\gamma_c^{\prime})^{-p};~~~~~~~~~~ \gamma^\prime_m \le \gamma^\prime \le \gamma^\prime_c\\
  (\gamma^\prime/\gamma_c^{\prime})^{-p-1}; ~~~~~~~ \gamma^\prime_c < \gamma^\prime \le \gamma^\prime_s
  \end{cases}
  ,
  \label{eq:e_dist_bw}
\end{eqnarray}
with $\gamma^\prime$ the Lorentz factor in the blastwave frame, and $p$ the electron index. Here
\begin{eqnarray} \label{eq:gam_m}
\gamma^{\prime}_m (t) = \left[\frac{m_p}{m_e} \epsilon_e \frac{p-2}{p-1} \Gamma_{\rm g}\right]; \, p>2,
\end{eqnarray}
is the minimum Lorentz factor, assuming a fraction $\epsilon_e$ of $u^\prime_g$ is acquired by the shock-accelerated electrons. The the cooling Lorentz factor is given by
\begin{eqnarray} \label{eq:gam_c}
\gamma^{\prime}_c (t) = \left[\frac{6 \pi m_e c^2 (1+z)}{\sigma_T c B{^{\prime}}^2(t) t \Gamma_{\rm g} (1+Y+\Gamma_{\rm g}^2u_{\rm ex}/u_B^\prime)}\right]
\end{eqnarray}
with $u_{\rm ex}$ the energy density of the external radiation field, and is found by equating the cooling time scale and dynamic time scale, and
\begin{eqnarray} \label{eq:gam_s}
\gamma^{\prime}_s (t) = \left[\frac{6 \pi e}{\phi \sigma_T B^{\prime}(t) (1+Y+\Gamma_{\rm g}^2u_{\rm ex}/u_B^\prime)}\right]^{1/2}
\end{eqnarray}
is the maximum or saturation Lorentz factor, found by equating the cooling time scale to the acceleration time scale. In the above equations, Eq.~\ref{eq:gam_m} is only valid when $\gamma^\prime_s(t)^{(p-2)}\gg\gamma^\prime_m(t)^{(p-2)}$, $m_p$ and $m_e$ is the proton and electron rest masses respectively, $\sigma_{T}$ the Thomson cross section, $c$ the speed of light, and $\phi \ge 1$ is the number of gyro-radius required for accelerating electrons to $\gamma^{\prime}_s$. The Compton $Y$-parameter in the slow-cooling, Thomson regime is given by
\begin{eqnarray}
Y = \sqrt{\epsilon_e/\epsilon_B} \times
\begin{cases}
  (t/t_0^{ssc})^{(2-p)/[2(4-p)]};~~~ {\rm Adiabatic-ISM} \\
  (t/t_0^{ssc})^{(2-p)/(4-p)};~~~~~~ {\rm Adiabatic-Wind} \\
  \end{cases}
  ,
  \label{eq:Y_slow}
\end{eqnarray}
with $t_0$ the transition time from the fast- to slow-cooling spectra, and $Y\approx \sqrt{\epsilon_e/\epsilon_B}\approx 1$ (implying $\epsilon_e \approx \epsilon_B$; see Equation (13) in \citet{Joshi2021}). Note that the second part of the equation is rather small for $p\sim2.0-2.3$ \citep{2001ApJ548787S}. $Y \approx 1$ value is imposed when we investigate the EC emission and need to reduce the SSC flux.

The normalisation constant $k$ in Equation~(\ref{eq:e_dist_bw}) can be found from the integrals $\int d\Omega \int_{\gamma^\prime_m}^{\gamma^\prime_s} n^\prime(\gamma^\prime) d\gamma= 4 n(R)\Gamma_g$, where $n(R)$ is the particle density in the surrounding environment in the lab frame, as
\begin{eqnarray} \label{eq:knorm}
    k = \frac{4n(R)\Gamma_{\rm g}}{\int_{\gamma^\prime_m}^{\gamma^\prime_c} (\gamma^\prime/\gamma_c^{\prime})^{-p} d\gamma^\prime + \int_{\gamma^\prime_c}^{\gamma^\prime_s} (\gamma^\prime/\gamma_c^{\prime})^{-p-1} d\gamma^\prime }.
\end{eqnarray}
Using the transformation $\gamma/\gamma^\prime=\Gamma_{\rm g}$ and $n(\gamma,\mu)=n^\prime(\gamma^\prime)(\gamma/\gamma^\prime)^2$ (with $\gamma$ the Lorentz factor in the lab frame, $\mu=\cos{\theta}$, and $\theta$ the angle of the blastwave with respect to the observer's line of sight), one obtains the lab frame electron distribution as
\begin{eqnarray}\label{eq:e_dist_bw_lab}
N(\gamma,\mu) &=& n^\prime (\gamma^\prime) \Gamma_{\rm g}^3 V \nonumber \\
 &=& \frac{k}{4\pi} \Gamma_g^3 V \times
\begin{cases}
  (\gamma^\prime/\gamma_c^{\prime})^{-p};~~~~~~~~~~ \gamma^\prime_m \le \gamma^\prime \le \gamma^\prime_c\\
  (\gamma^\prime/\gamma_c^{\prime})^{-p-1}; ~~~~~~~ \gamma^\prime_c < \gamma^\prime \le \gamma^\prime_s \,.
  \end{cases}
\end{eqnarray}
Here $V=(4/3)\pi R^3$ is the volume of the blastwave in the lab frame.

\subsection{External Compton radiation} \label{sec:ECrad}

Electrons can lose energy by Compton scattering photons of an external radiation field. This external Compton (EC) process is relevant when the energy density of the external radiation field in the jet frame exceeds that of the energy density in the magnetic field, i.e., $u_{\rm ex} \Gamma_g^2 \gtrsim u^\prime_B = 4n(R)m_p c^2\Gamma_g^2 \epsilon_B$. This gives a condition on the number density of particles in the environment as $n(R) \lesssim u_{\rm ex}/(4m_pc^2\epsilon_B)$ for EC to be important. For CMB with energy density $4.17\times 10^{-13}(1+z)^4$ erg/cm$^3$, the requirement is then 
\begin{equation}\label{eq:nR}
    n(R) \lesssim 7\times 10^{-5} (1+z)^4\epsilon_B^{-1} ~{\rm cm}^{-3} \,.
\end{equation}
For a wind-type environment this condition leads to a time scale of
\begin{equation}
    t_{\rm EC, wind} \gtrsim 5.7\times 10^5\,  \epsilon_{B,-2} A_{\star, -2}^2 E_{54}^{-1} \, {\rm s},
\end{equation}
when VHE emission by EC is important. Here $\epsilon_{B,-2} = \epsilon_{B}/0.01$, $A_{\star, -2} = A_{\star}/0.01$ and $E_{54} = E_k/10^{54}$~erg.

To calculate the EC spectrum we followed the same numerical calculations as given in \citet{GKM2001}, to study the blastwave evolution (Section~\ref{sec:blastwave}) and the particle transport (Section~\ref{sec:transport}) of GRB afterglow emission. This model assumes an isotropic distribution of soft mono-energetic photons upscattered by non-thermal electrons described by a power-law density distribution. We adapted this model by describing the electron density as a broken power-law (see Equation~(\ref{eq:e_dist_bw})), assuming that non-thermal ultra-relativistic electrons in the blastwave upscatter low-energy photons from an external radiation field via IC.

In the Klein-Nishina (KN) and Thomson limits we obtain the specific luminosity $L$ per energy $\epsilon$ interval per solid angle $\Omega$ in the same manner as \cite{GKM2001}, using the electron distribution in Equation~(\ref{eq:e_dist_bw_lab}), scattering rate of the photons $dN_p/dtd\epsilon=3\sigma_T c f(x)/4\epsilon_0\gamma^2$, observed photon energy $\epsilon$ (in units of $m_ec^2$), the photon number density $n_p={u_{\rm ex}}/\epsilon_0m_ec^2$, the photon energy density $u_{\rm ex}$ and the target photon energy $\epsilon_0$ (of the external radiation field), the expression yields
\begin{equation} \label{eq:lum}
\begin{split}
\epsilon\frac{dL}{d\epsilon d\Omega} & = D^{3+p} \frac{3kV\sigma_{\rm T}c{u_{\rm ex}}}{16\pi} \left(\frac{\epsilon}{\epsilon_0}\right)^2 \\
 & \times 
  \begin{cases}
  \int_{\gamma_{\rm m}}^{\gamma_{\rm c}} \gamma^{-(2+p)}\gamma_{c}^{p}f(x)d\gamma; ~~~~~~~~~~~~ \gamma_m \le \gamma \le \gamma_c \\
  \int_{\gamma_{\rm c}}^{\gamma_{\rm s}} \gamma^{-(3+p)}\gamma_{c}^{(1+p)}f(x)d\gamma; ~~~~~~ \gamma_c < \gamma \le \gamma_s,
  \end{cases}
\end{split}
\end{equation}
in erg-s$^{-1}$, and assuming $\Gamma_{\rm g}\approx D$ with $D$ being the Doppler factor, and $k$ the normalisation factor in Equation~(\ref{eq:knorm}). In the Klein-Nishina case, for energies $\epsilon_{\rm min,KN}\leq\epsilon\leq\epsilon_{\rm max,KN}$, the lower limit of integration in Equation~(\ref{eq:lum}) is found by setting $x=1$, and is given by
\begin{eqnarray} \label{eq:gammin}
\gamma_{\rm min}&=&\frac{\epsilon\epsilon_0+\sqrt{\epsilon^2\epsilon_0^2 + \epsilon\epsilon_0}}{2\epsilon_0},
\end{eqnarray}
when $\gamma_{\rm min}>\gamma_m$ we used $\gamma_{\rm min}$, and vice versa.

Additionally, we considered the effect of the extragalactic background light (EBL) $\gamma$-ray attenuation on our EC model. Thus, we fit an attenuated model of the form
\begin{eqnarray} \label{eq:ebl}
\frac{dN}{dE} = \left(\frac{dN}{dE}\right)_{\rm EC}e^{-\tau(E,z)} = \left(\frac{L}{4\pi D_{\rm L}^2}\right)_{\rm EC}e^{-\tau(E,z)},
\end{eqnarray}
where $(dN/dE)_{\rm EC}$ is the intrinsic EC energy flux (calculated by dividing Equation~(\ref{eq:lum}) by $4\pi D_{\rm L}^2$, with $D_{\rm L}$ the luminosity distance of the source), the exponential term corresponding to the attenuation, $\tau$ is the energy-dependent optical depth for a source at redshift $z$.

\subsection{Synchrotron and Synchrotron self-Compton radiation} 
\label{sec:SSCrad}

In addition to the external Compton described above electrons will also emit radiation due to the synchrotron and SSC processes. To estimate these radiation spectra we have followed the formalism as discussed in \citet{1998ApJ...497L..17S, 2001ApJ548787S, Joshi2021}. For the GRBs discussed here, the late emission occurs in the slow cooling regime for which the synchrotron spectrum can be calculated using
%
\begin{align}
 F_{\nu, \rm{slow}} &= f_{\nu, \rm{max}}
\begin{cases}
  (\frac{\nu}{\nu_m})^{1/3};~~~~~~~~~~~~~~~~~~~~~~~ \nu \le \nu_m\\
  (\frac{\nu}{\nu_m})^{-(p-1)/2};~~~~~~~~~~~~~~ \nu_m < \nu < \nu_c\\
    (\frac{\nu_c}{\nu_m})^{-(p-1)/2} (\frac{\nu}{\nu_c})^{-p/2};~~~~~~ \nu \ge \nu_c \,.
  \end{cases}
  \label{slow_spectrum}
\end{align}
%
Here $\nu$ is the synchrotron radiation frequency and $\nu_m$ is the synchrotron frequency corresponding to the minimum electron Lorentz factor $\gamma^\prime_m$, $\nu_c$ corresponds to the cooling Lorentz factor $\gamma^\prime_c$ and $\nu_s$ corresponds to the maximum value of the electron Lorentz factor $\gamma^\prime_s$. The value of the maximum synchrotron flux density $f_{\nu, \rm{max}}$ is taken from \citet{2013PhRvD..88j3003R}. During the SSC scattering the synchrotron photons are upscattered by the non-thermal electrons and in the Thomson regime the upscattered frequencies are $\nu_m^{\rm ssc} \simeq 2 \gamma_m^{\prime 2} \nu_a, \nu_c^{\rm ssc} \simeq 2 \gamma_c^{\prime 2} \nu_c$. The SSC spectra have been calculated using the smooth approximation, where for each electron energy the possible interactions with the synchrotron photons are integrated, as discussed by \citet{2001ApJ548787S}. However, KN effects will start above Lorentz factor $\gamma^\prime_{\rm KN} \simeq m_e c^2 \Gamma_g/h \nu_c (1+z)$ \citep{Joshi2021} and corresponding maximum photon energy in the Thomson regime for the wind case would be $E_{\gamma, \rm cut}^{\rm ssc, Th} = 5.5~ {\rm TeV} A_{\star, -2}^{3/2} \epsilon_{B, -1}^{3/2} t_2^{-1} (1 + Y)^2$. Above this energy the SSC spectrum is steeper due to the KN effects as discussed in \citet{2009ApJ_Nakar_KN} and the same is applied to our SSC spectrum.

\section{Multiwavelength modelling of GRBs with late time VHE Emission}
\label{sec:results}

We model optical to VHE gamma-ray data from the afterglows of GRB~180720B and GRB~190829A detected by the H.E.S.S. in this section. We have considered synchrotron radiation (SR), SSC and EC from relativistic electrons accelerated in the blastwaves of these GRBs. Attenuation of VHE photons in the EBL is included in our modelling. We have considered both the ISM and wind environments but the wind environment gives a better fit to data. We also considered infrared, optical and CMB as the external radiation field, but the CMB gives a better fit to the VHE data. Below, we have discussed the observational properties and simultaneous interpretation of their light curves and SEDs.

\subsection{GRB~180720B} \label{sec:grb180720B}

\begin{figure}
    \includegraphics[width=0.95\columnwidth]{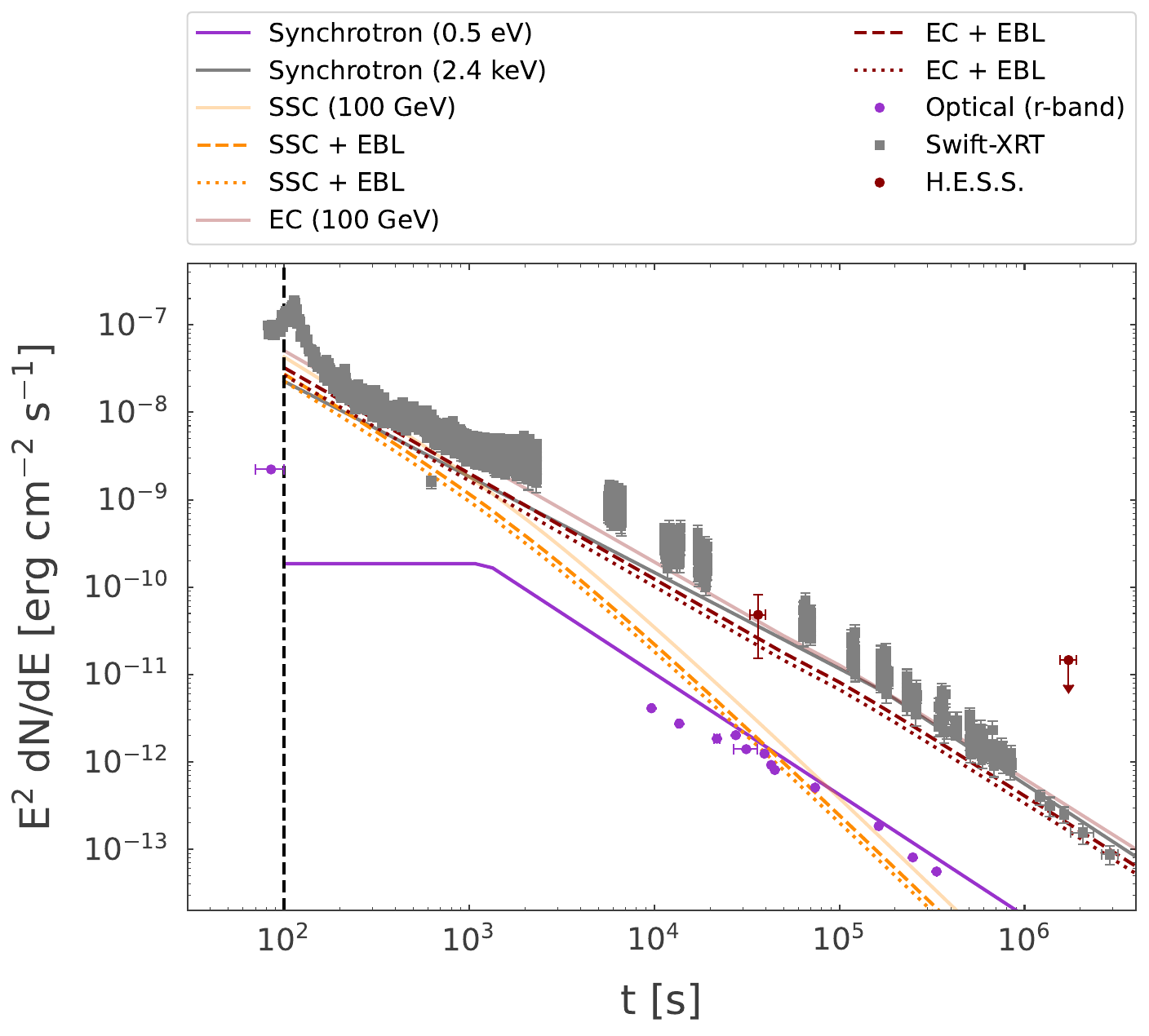}
    \caption{Multiwavelength light curves including optical (purple; r-band), the \emph{Swift}-BAT spectra (15 keV-150 keV) extrapolated to the XRT band (0.3–10 keV) for a combined light curve (grey; \url{http://www.swift.ac.uk/xrt_curves}), and \emph{H.E.S.S.} (100-440~GeV; \citealt{Abdalla2019_GRB180720b}) observations, for long GRB~180720B in the wind case for the CMB target photons ($\epsilon_0=8.9\times 10^{-4}$~eV). We indicate $t_{\rm dec,w}$ (black dashed line), and show the synchrotron (purple and grey lines), SSC (orange lines) and EC (darkred lines). The dashed \citep{Finke2010} and dotted \citep{Inoue2013} lines take into account the EBL attenuation at energies $>100$~GeV. See relevant physical parameters listed in Table~\ref{tab:fit_values}.}
\label{fig:grb180720b_wind_lc}
\end{figure}

\begin{figure}
    \includegraphics[width=0.95\columnwidth]{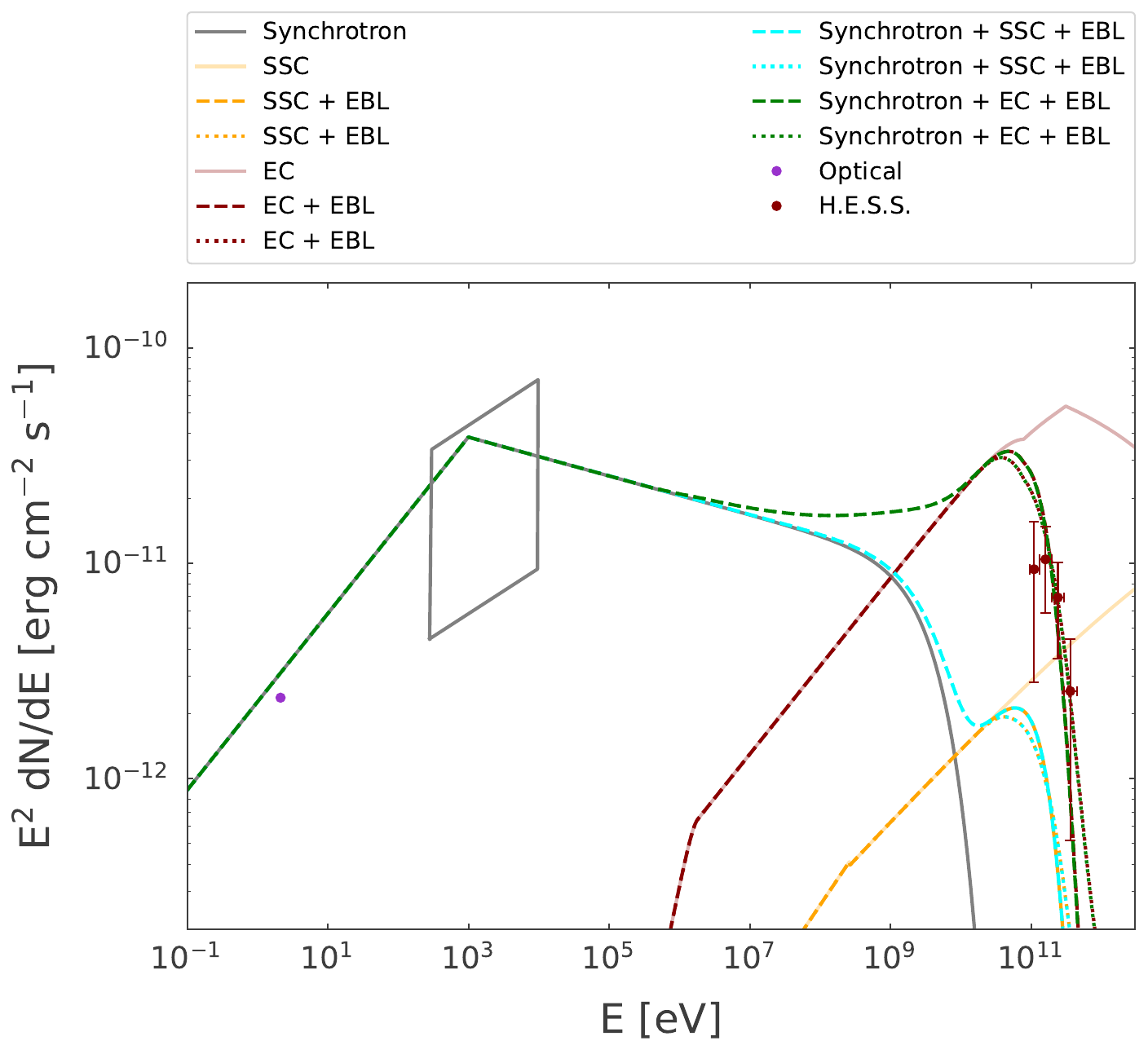}
    \caption{Multiwavelength SED of GRB~180720B corresponding to the light curves in Figure~\ref{fig:grb180720b_wind_lc}. The grey box shows the spectrum and uncertainty of the X-ray data (an extrapolation to the time window of H.E.S.S. observation), and the purple point is the optical data \citep{Fraija2019}. The cyan and green lines are for the joint spectral fits of the synchrotron with SSC and EC respectively. The SSC and EC components are corrected for EBL attenuation using the models of \citet{Finke2010} (dashed lines) and \citet{Inoue2013} (dotted lines).}
\label{fig:grb180720b_wind_SED}
\end{figure}

GRB 180720B, with a redshift $z$=0.653 and a $T_{90}=48.9\pm 0.4$~s, is one of the brightest VHE sources and was detected by H.E.S.S. at energies up to 440~GeV in the late afterglow at $T_0+10$~hours \citep{Abdalla2019_GRB180720b}. During its prompt phase, the isotropic energy released in gamma-rays is $(6 \pm 0.1) \times 10^{53}$ erg \citep{Abdalla2019_GRB180720b}. The gamma-ray observations by Fermi-LAT also infer a smooth transition from prompt to afterglow \citep{2020A&A...636A..55R}. The late time VHE emission in this object also implies that the blastwave has slowed down considerably.

Figure~\ref{fig:grb180720b_wind_lc} represents the multi-wavelength light curves for this source from optical (r-band), X-ray, to VHE. We focused on the explanation for the late-time emission of each band at $10^4<t<10^{6.7}$~s, corresponding to the late-time afterglow range of the H.E.S.S. observations. In this time range a set of values are chosen such that the requirement in Equation~(\ref{eq:nR}) and $Y\approx1$ are satisfied, as well as to fit each wave band simultaneously, with the exception of the photon energy. These values are: $E_{k}=4.7\times 10^{54}$ erg, $\Gamma$=318, $t_{\rm dec,w}=100$~s, $A_{\rm \star}=1\times 10^{-1}$, $p$=2.18, $\epsilon_e=3\times 10^{-2}$, $\epsilon_B=3\times 10^{-2}$ (with the latter two values chosen such that $Y\approx1$ is fixed), and $\phi$=10 (see Table~\ref{tab:fit_values}). These parameters are used to determine the quantities as described in Section~\ref{sec:blastwave} and~\ref{sec:transport} to calculate the luminosity and therefore the EC energy flux. The optical data are described by SR at energies $0.5$~eV, and the \emph{Swift}-XRT data by SR at energies $2.4$~keV (data from figure~(1) in \citet{Abdalla2019_GRB180720b}). 
The set of parameters result in an estimation for $t_{\rm dec,w}$ that excludes the early time data points. We have chosen the parameter space as such that we can compare the EC emission to the H.E.S.S. data at energies $100$~GeV. We included the SSC emission as well since this has been used primarily to explain the VHE observations, however, for our chosen set of model parameters the SSC energy flux is expected to be significantly lower \citep{Joshi2021}. We applied Equation~(\ref{eq:ebl}) to both the SSC and EC since both these emission mechanisms can fit the data at VHE energies, considering only the EBL models of \citet{Finke2010} and \citet{Inoue2013} with the former model estimating the attenuation slightly larger than the latter one. In the ISM case (not shown), the fit to the optical and X-ray data requires a steeper slope, and the slope of the EC fit is too steep. Therefore we could not find a set of parameters to fit the observations in the different wavebands simultaneously, including the upper limit of the H.E.S.S. data, compared to the wind case. We gave preference to X-ray over the optical data.

Figure~\ref{fig:grb180720b_wind_SED} shows the energy spectrum associated with Figure~\ref{fig:grb180720b_wind_lc}, using the same physical parameters at a $t=3.6\times 10^4$~s ($T_0+10$~hrs). This set of parameters result in a good prediction of the EC fit to the VHE data, as well as the synchrotron to the X-ray and optical \citep{Fraija2019}. We included a joint fit of SR with both SSC and EC respectively, although the contribution of SSC to the joint fit will be $10\%$ which is negligible compared to that of the EC for the chosen values to explain the VHE observations via EC. In the ISM case the SED, for these same parameter values as for the light curves in the same case, does not produce a reasonable fit through the X-ray data, with the SSC flux relatively higher than in the wind case. However, the EC emission still result in a good fit to the H.E.S.S. observations. In both cases the EBL model of \citet{Finke2010} results in a slightly larger flux, however the model of \citet{Inoue2013} results in a slightly extended energy tail.

It is important to note that the energy flux of the H.E.S.S. observations in Figure~\ref{fig:grb180720b_wind_lc} is slightly higher than those in Figure~\ref{fig:grb180720b_wind_SED}. This is due to the different methods used to do the light curve and spectral model fitting compatible with the instrument. This is also the case for the observations of GRB~190829A discussed in the next section.

\begin{figure}
	\includegraphics[width=0.95\columnwidth]{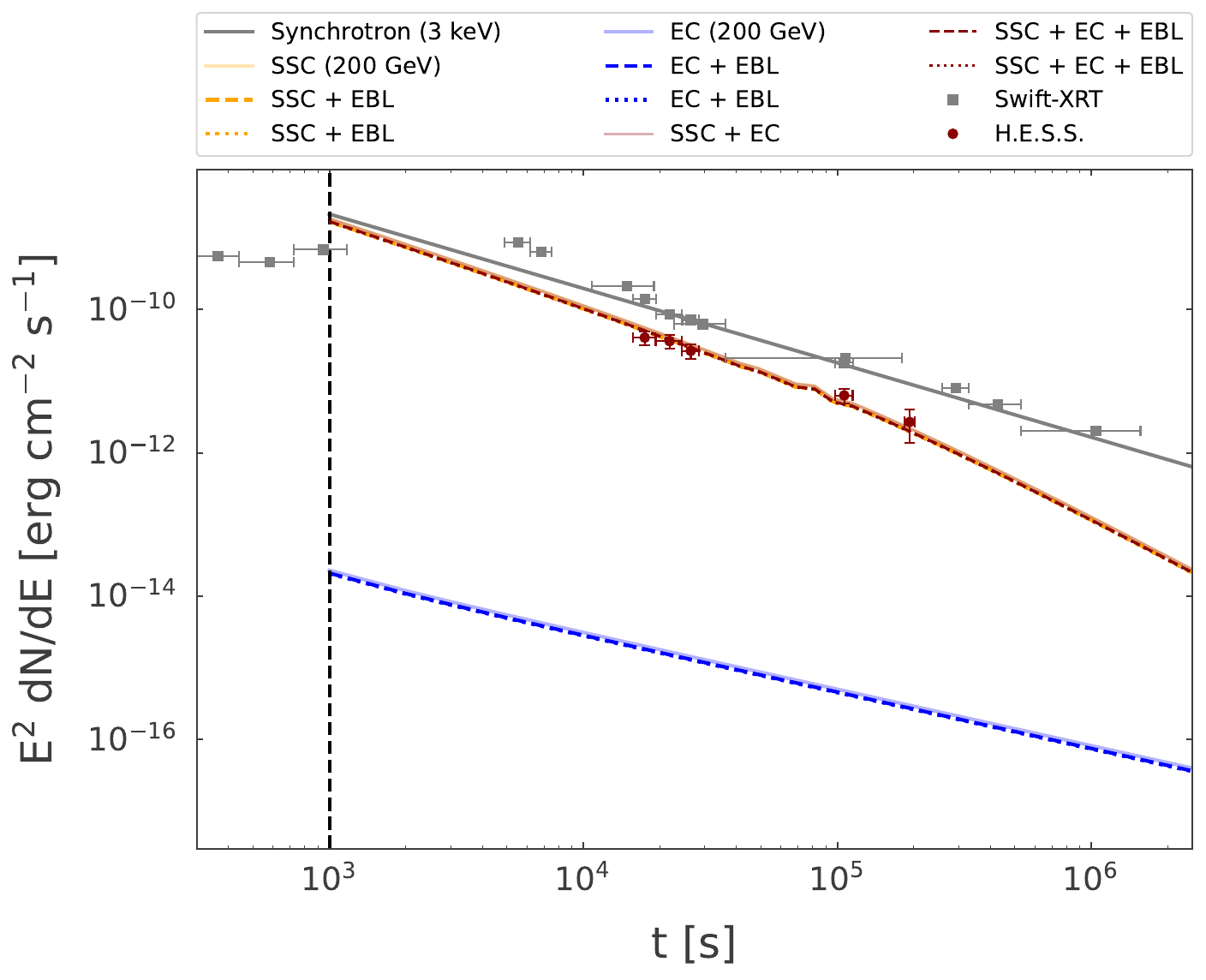}
	\caption{Same as Figure~\ref{fig:grb180720b_wind_lc} for long GRB~190829A, without optical observations. We include observations for more than one night. These include the temporal evolution of the energy flux detected in X-rays by \emph{Swift}-XRT (grey squares), including those that were simultaneous with the \emph{H.E.S.S.} observations (200~GeV-4~TeV; \citealt{Abdalla2021_GRB190829a}), with an additional night 3 for the latter. In this case the combined fit of EC (blue) with SSC (orange) is indicated by the dark red lines.}
    \label{fig:grb190829a_wind_lc}
\end{figure}

\begin{figure}
	\includegraphics[width=0.95\columnwidth]{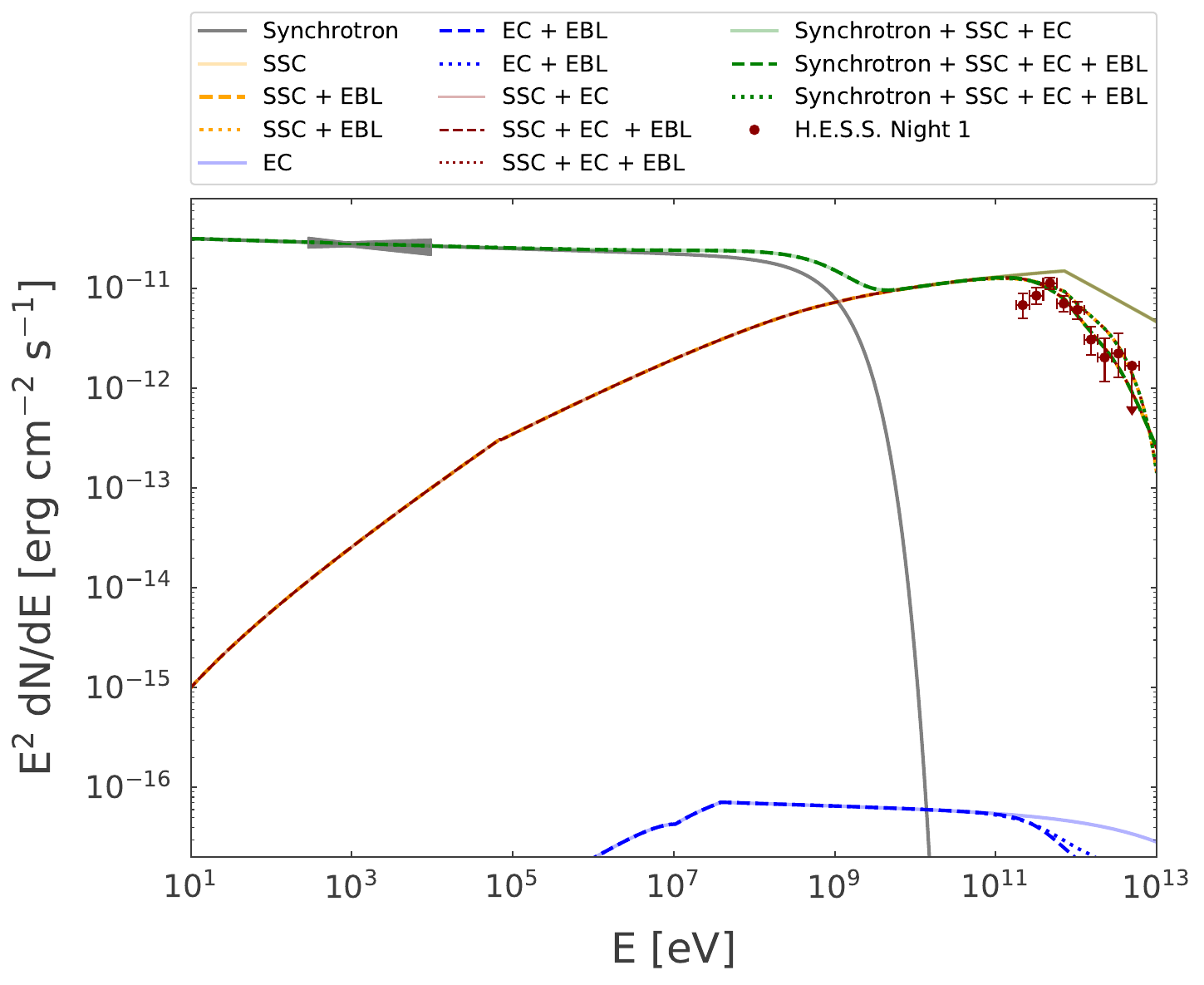}
	\includegraphics[width=0.95\columnwidth]{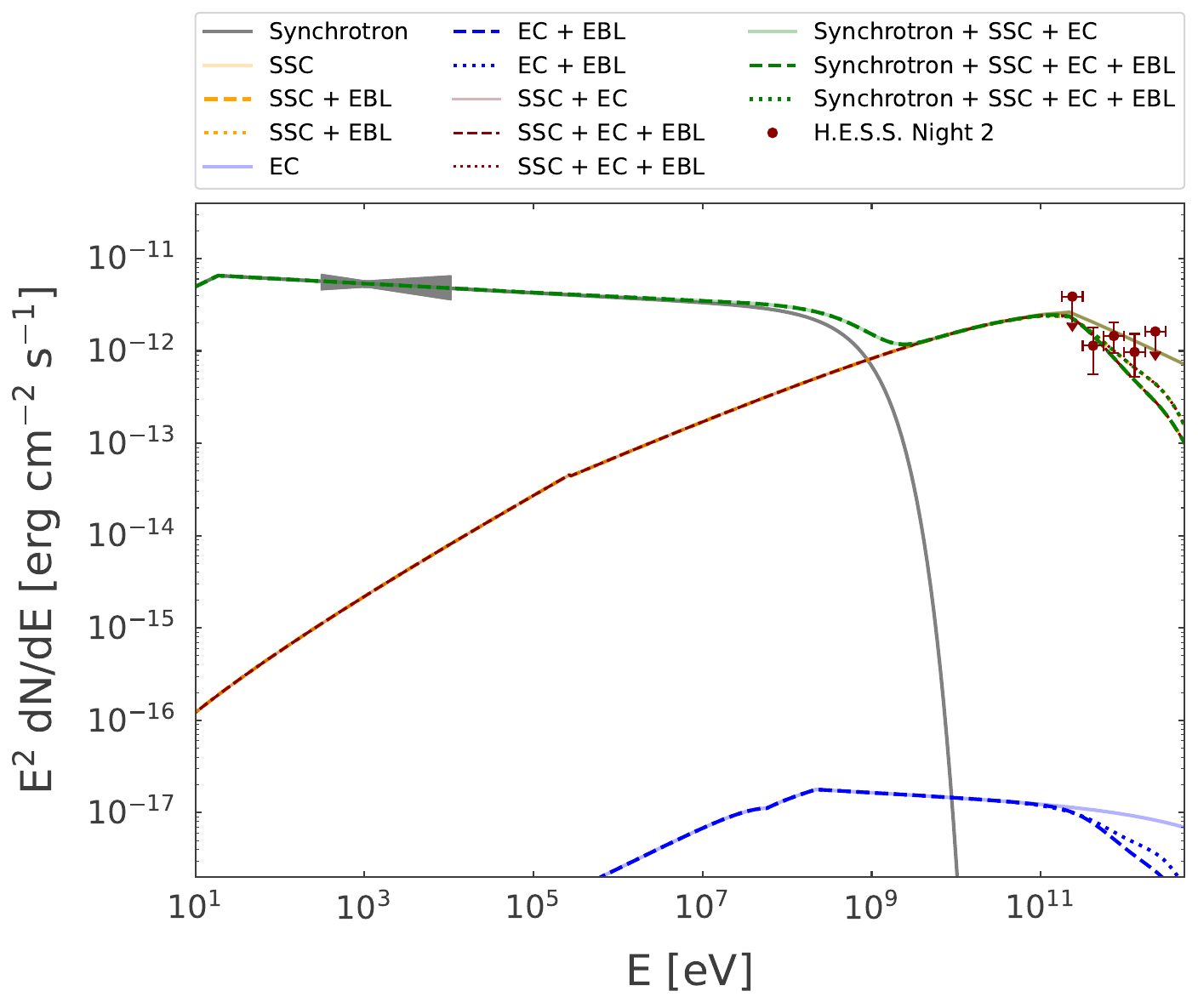}
    \caption{Same as Figure~\ref{fig:grb180720b_wind_SED}, we show the SED for night 1 (top) and night 2 (bottom), with the signal for night 3 very low, for long GRB~190829A without optical observations. The shaded grey regions show the spectrum and uncertainty of the \emph{Swift}-XRT data \citep{Abdalla2021_GRB190829a}, and the green lines indicate the joint spectral fits of the synchrotron (grey) with the combined SSC (orange) and EC (blue) components, indicated by the dark red lines.}
    \label{fig:grb190829a_wind_SED}
\end{figure}

\subsection{GRB~190829A}
\label{sec:grb190829A}

H.E.S.S. also detected VHE radiation from GRB 190829A, with a $z=0.0785$ and a T$_{90}=63$~s, above 3~TeV for the first night at $T_0+4.3$~hr, above 1~TeV for the second night at $T_0+27.2$~hr \citep{Abdalla2021_GRB190829a}, and for the third night no significant signal was detected. 
In the afterglow emission, the X-ray and GeV-TeV (0.18 to 3.3 TeV) light curves follow similar decay features \citep{Abdalla2021_GRB190829a}.

In Figure~\ref{fig:grb190829a_wind_lc} we present the multiwavelength light curves for this source in X-ray and VHE. We consider the late-time emission of each band 
corresponding to the H.E.S.S. observations for the afterglow in nights 1-3. In this time range a set of values are chosen such that the requirement in Equation~(\ref{eq:nR}) and $Y\approx1$ are satisfied as well as to fit each wave band simultaneously, with the exception of the photon energy. These values are: $E_k=2\times 10^{52}$~erg, $\Gamma=42$, $t_{\rm dec,w}=1\times 10^{3}$~s, $A_{\rm \star}=9\times 10^{-2}$, $p=2.073$, $\epsilon_e=9.1\times 10^{-2}$, $\epsilon_B=9.1\times 10^{-2}$ (with the latter two values chosen such that $Y\approx1$ is fixed), and $\phi=10$ (see Table~\ref{tab:fit_values}). The \emph{Swift}-XRT data are taken from \citet{Abdalla2021_GRB190829a} and include observations from night 1 and night 2, and are described by SR at energies $3$~keV. There were no coinciding observations for night 3 in the X-ray band. 
For this source a combined fit of the SSC and EC components explains the H.E.S.S. data (including all three nights) at energies $200$~GeV and found a reasonable fit to explain the VHE observations. This is due to the strict X-ray data constraint. However, for the chosen parameter space the SSC energy flux dominates the EC \citep{Joshi2021}. We applied Equation~(\ref{eq:ebl}) to both the SSC and EC since we fit data at VHE energies, considering only the EBL models of \citet{Finke2010} and \citet{Inoue2013} with both models estimating the attenuation at the same flux level. In the ISM case (not shown), the fit to the X-ray data requires a steeper slope, thereby missing the second and third night. Therefore we could not find a set of parameters to fit the observations in the different wavebands simultaneously, compared to the wind case.

Figure~\ref{fig:grb190829a_wind_SED} shows the energy spectrum for night 1 (top panel) and night 2 (bottom panel) associated with Figure~\ref{fig:grb180720b_wind_lc}, using the same physical parameters but with $E_k=5.6\times 10^{51}$~erg, $p=2.05$, and $t_{n1}=1.55\times 10^4$~s ($T_0+4.3$~hrs) for night 1, and $E_k=5\times 10^{51}$~erg, $p=2.1$, and $t_{n2}=9.79\times10^4$~s ($T_0+27.2$~hrs) for night 2 (see Table~\ref{tab:fit_values} column Night 1 and Night 2, respectively). We varied the $E_k$ and $p$ to fit the spectral data for each night, however the $E_k$ value chosen to fit the light curve is an average of the values chosen for the spectra of both nights. This resulted in a good prediction of a combination of the SSC and EC components to the VHE data, as well as the synchrotron to the X-ray data.  
We included a fit of SR with the combined SSC and EC components, although the contribution of the EC flux to the combined fit is negligible compared to the contribution of the SSC. In the ISM case (not shown) the SED, for these same parameter values as for the light curves in the same case, does not produce a reasonable fit through the VHE data, with the SSC spectrum not as hard as in the wind case to fit the spectral observations. However, the SSC and EC combination still results in a good fit to the H.E.S.S. observations.
In both cases of the EBL model of \citet{Inoue2013} result in a slightly larger flux, however the model of \citet{Finke2010} result in a slightly extended energy tail.

\section{DISCUSSIONS AND CONCLUSIONS} \label{sec:discussion}

\begin{table}
\centering
\caption{
Summary of the afterglow parameter choices (with $Y$ a fixed value) from fitting the multiwavelength light curves and spectra simultaneously.
}
\label{tab:fit_values}
\begin{tabular}{lcccc}
    \hline
    Parameter & GRB~180720B & \multicolumn{3}{c}{GRB 190829A} \\
     & & Light curve & Night 1 & Night 2 \\
    \hline
    $E_{k}$ (erg) & $4.7\times 10^{54}$ & $2\times 10^{52}$ & $5.6\times 10^{51}$ & $5\times 10^{51}$ \\
	$\Gamma$ & 318 & 42 & 42 & 42 \\
    $t_{\rm dec,w}$ (s) & $100$ & $1\times 10^{3}$ & $1\times 10^{3}$ & $1\times 10^{3}$ \\
    $t$ (s) & $3.6\times 10^4$ & -- & $1.55\times 10^{4}$ & $9.79\times 10^{4}$ \\
    $A_{\rm \star}$ & $1\times 10^{-1}$ & $9\times 10^{-2}$ & $9\times 10^{-2}$ & $9\times 10^{-2}$ \\
    $p$ & 2.18 & 2.073 & 2.05 & 2.1 \\
    $\epsilon_e$ & $3\times 10^{-2}$ & $9.1\times 10^{-2}$ & $9.1\times 10^{-2}$ & $9.1\times 10^{-2}$ \\
	$\epsilon_B$ & $3\times 10^{-2}$ & $9.1\times 10^{-2}$ & $9.1\times 10^{-2}$ & $9.1\times 10^{-2}$ \\
    \hline
\end{tabular}
\label{tab:multicol}
\end{table}

In this work, we have shown that late-time VHE afterglow of GRB~180720B and 
GRB~190829A detected by H.E.S.S. can be produced by the interaction of relativistic electrons in the GRB blastwave with the CMB radiation. We have explained the light-curves in optical, X-ray, VHE and corresponding SED of these bursts using a synchrotron, SSC and EC radiation model. For GRB~180720B the contribution due to the SSC mechanism is negligible based on the model parameters used, but this is not the case for GRB~190829A which result in a combiantion fit of SSC and EC. In particular, we found that one needs a relatively low environment density, although not unusual, and $\epsilon_e \sim \epsilon_B$ for the EC emission to be dominant. However, even with $Y\approx1$ imposed the EC is not always dominant as shown for GRB~190829A where the EC flux is very low and SSC dominating the fit.
A summary of the model parameters are listed in Table~\ref{tab:fit_values}. 
For both GRBs the wind scenario was favoured over the ISM. Also, the EBL correction for the two different EBL models do not deviate too much. For GRB 190829A we used parameters that fit the light curves and SEDs simultaneously, but $E_k$ and $p$ of the blastwave for both varies slightly due to the spectral shape for night 1 being harder than night 2. In the light curve we tried to fit three nights at once, again resulting in a slightly different $E_k$ and $p$. Thus, fitting GRB 190829A is more complex since the observations are spread over three nights and with limited observations available for GRB~180720B and GRB~190829A one can not get a better fit of the model to the data. The fit of X-ray is preferred over optical since the host galaxy contributes the vast majority of the optical emission. 

Our EC modelling of VHE emission differs from the SSC model used for fitting VHE data from GRB~180720B where the blastwave traverses in a constant density ISM \citep{2019ApJ...884..117W,Fraija2019}. The isotropic energy released in the gamma rays from GRB~190829A during the prompt emission was $\sim 10^{50}$ erg \citep{2020ApJ...898...42C} and termed this as an example of low-luminosity (LL) GRB. We found that indeed the kinetic energy of the burst is lower than that of GRB~180720B by more than an order of magnitude. SSC models were also used to explain the VHE gamma-ray emission from GRB~190829A \citep{2021ApJ...917...95Z, Salafia2022} and its low luminosity was also used as a proxy for off-axis VHE emission \citep{2021MNRAS.504.5647S}. Further, EC models with target photons from the early-afterglow flare was used to explain VHE radiation \citep{2021ApJ92055Z}. A photo-hadronic model was also used to produce VHE gamma rays in this burst \citep{2022ApJ92970S}.

In summary, we found that inverse-Compton scattering of CMB photons by relativistic electrons in the GRB blastwave can explain VHE emission detected for some GRBs at late time ($\sim$ hours) when the energy density in the CMB radiation becomes comparable to that of the magnetic energy density in the blastwave. We found that the energy densities in the magnetic field and in relativistic electrons are both of the order of a few percent of the blastwave kinetic energy.

\section*{Acknowledgements}
We thank the H.E.S.S. collaboration for sharing the data for the GRBs studied in this paper and Maria Petropolou for discussion. We also thank the MNRAS Referee for constructive suggestions that improved the paper. The research of MB was supported by a GES fellowship of University of Johannesburg, where this work was completed. SR acknowledges support from the National Research Foundation, South Africa for a BRICS STI grant and other grants from the University of Johannesburg Research Council. JCJ was partially supported by a grant from the University of Johannesburg Research Council to visit CAPP.

\section*{Data Availability}
All data used in this article are publicly available from the H.E.S.S. Collaboration (\url{https://www.mpi-hd.mpg.de/HESS/}) and from the \emph{Swift}-XRT website: \url{https://www.swift.ac.uk/analysis/xrt/}.



\bibliographystyle{mnras}
\bibliography{biblio} 

\begin{thebibliography}{}
\makeatletter
\relax
\def\mn@urlcharsother{\let\do\@makeother \do\$\do\&\do\#\do\^\do\_\do\%\do\~}
\def\mn@doi{\begingroup\mn@urlcharsother \@ifnextchar [ {\mn@doi@}
  {\mn@doi@[]}}
\def\mn@doi@[#1]#2{\def\@tempa{#1}\ifx\@tempa\@empty \href
  {http://dx.doi.org/#2} {doi:#2}\else \href {http://dx.doi.org/#2} {#1}\fi
  \endgroup}
\def\mn@eprint#1#2{\mn@eprint@#1:#2::\@nil}
\def\mn@eprint@arXiv#1{\href {http://arxiv.org/abs/#1} {{\tt arXiv:#1}}}
\def\mn@eprint@dblp#1{\href {http://dblp.uni-trier.de/rec/bibtex/#1.xml}
  {dblp:#1}}
\def\mn@eprint@#1:#2:#3:#4\@nil{\def\@tempa {#1}\def\@tempb {#2}\def\@tempc
  {#3}\ifx \@tempc \@empty \let \@tempc \@tempb \let \@tempb \@tempa \fi \ifx
  \@tempb \@empty \def\@tempb {arXiv}\fi \@ifundefined
  {mn@eprint@\@tempb}{\@tempb:\@tempc}{\expandafter \expandafter \csname
  mn@eprint@\@tempb\endcsname \expandafter{\@tempc}}}

\bibitem[\protect\citeauthoryear{{Abbasi} et~al.,}{{Abbasi}
  et~al.}{2023}]{2023ApJ94626A}
{Abbasi} R.,  et~al., 2023, \mn@doi [\apjl] {10.3847/2041-8213/acc077}, \href
  {https://ui.adsabs.harvard.edu/abs/2023ApJ...946L..26A} {946, L26}

\bibitem[\protect\citeauthoryear{{Abdalla} et~al.,}{{Abdalla}
  et~al.}{2019}]{Abdalla2019_GRB180720b}
{Abdalla} H.,  et~al., 2019, \nat, 575, 464

\bibitem[\protect\citeauthoryear{{Abdalla} et~al.,}{{Abdalla}
  et~al.}{2021}]{Abdalla2021_GRB190829a}
{Abdalla} H.,  et~al., 2021, Science, 372, 1081

\bibitem[\protect\citeauthoryear{{Alves Batista}}{{Alves
  Batista}}{2022}]{Batista2022arXiv221012855A}
{Alves Batista} R.,  2022, arXiv e-prints, \href
  {https://ui.adsabs.harvard.edu/abs/2022arXiv221012855A} {p. arXiv:2210.12855}

\bibitem[\protect\citeauthoryear{{Asano}, {Inoue}  \&
  {M{\'e}sz{\'a}ros}}{{Asano} et~al.}{2009}]{2009ApJ...699..953A}
{Asano} K.,  {Inoue} S.,   {M{\'e}sz{\'a}ros} P.,  2009, \mn@doi [\apj]
  {10.1088/0004-637X/699/2/953}, \href
  {https://ui.adsabs.harvard.edu/abs/2009ApJ...699..953A} {699, 953}

\bibitem[\protect\citeauthoryear{{Baktash}, {Horns}  \& {Meyer}}{{Baktash}
  et~al.}{2022}]{baktash22}
{Baktash} A.,  {Horns} D.,   {Meyer} M.,  2022, arXiv e-prints, \href
  {https://ui.adsabs.harvard.edu/abs/2022arXiv221007172B} {p. arXiv:2210.07172}

\bibitem[\protect\citeauthoryear{{Barniol Duran} \& {Kumar}}{{Barniol Duran} \&
  {Kumar}}{2011}]{2011MNRAS.412..522B}
{Barniol Duran} R.,  {Kumar} P.,  2011, \mn@doi [\mnras]
  {10.1111/j.1365-2966.2010.17927.x}, \href
  {https://ui.adsabs.harvard.edu/abs/2011MNRAS.412..522B} {412, 522}

\bibitem[\protect\citeauthoryear{{Blandford} \& {McKee}}{{Blandford} \&
  {McKee}}{1976}]{blandford_mckee_soln_76}
{Blandford} R.~D.,  {McKee} C.~F.,  1976, \mn@doi [Physics of Fluids]
  {10.1063/1.861619}, \href {http://adsabs.harvard.edu/abs/1976PhFl...19.1130B}
  {19, 1130}

\bibitem[\protect\citeauthoryear{{B{\"o}ttcher} \& {Dermer}}{{B{\"o}ttcher} \&
  {Dermer}}{1998}]{1998ApJ...499L.131B}
{B{\"o}ttcher} M.,  {Dermer} C.~D.,  1998, \mn@doi [\apjl] {10.1086/311366},
  \href {https://ui.adsabs.harvard.edu/abs/1998ApJ...499L.131B} {499, L131}

\bibitem[\protect\citeauthoryear{{Chand} et~al.,}{{Chand}
  et~al.}{2020}]{2020ApJ...898...42C}
{Chand} V.,  et~al., 2020, \mn@doi [\apj] {10.3847/1538-4357/ab9606}, \href
  {https://ui.adsabs.harvard.edu/abs/2020ApJ...898...42C} {898, 42}

\bibitem[\protect\citeauthoryear{{Cheung}}{{Cheung}}{2022}]{Cheung2022arXiv221014178C}
{Cheung} K.,  2022, arXiv e-prints, \href
  {https://ui.adsabs.harvard.edu/abs/2022arXiv221014178C} {p. arXiv:2210.14178}

\bibitem[\protect\citeauthoryear{{Chiang} \& {Dermer}}{{Chiang} \&
  {Dermer}}{1999}]{1999ApJ699C}
{Chiang} J.,  {Dermer} C.~D.,  1999, \mn@doi [\apj] {10.1086/306789}, \href
  {https://ui.adsabs.harvard.edu/abs/1999ApJ...512..699C} {512, 699}

\bibitem[\protect\citeauthoryear{{Da Vela}, {Mart{\'\i}-Devesa}, {Saturni},
  {Veres}, {Stamerra}  \& {Longo}}{{Da Vela}
  et~al.}{2023}]{2023PhRvD.107f3030D}
{Da Vela} P.,  {Mart{\'\i}-Devesa} G.,  {Saturni} F.~G.,  {Veres} P.,
  {Stamerra} A.,   {Longo} F.,  2023, \mn@doi [\prd]
  {10.1103/PhysRevD.107.063030}, \href
  {https://ui.adsabs.harvard.edu/abs/2023PhRvD.107f3030D} {107, 063030}

\bibitem[\protect\citeauthoryear{{Das} \& {Razzaque}}{{Das} \&
  {Razzaque}}{2023}]{2023AnA...670L..12D}
{Das} S.,  {Razzaque} S.,  2023, \mn@doi [\aap] {10.1051/0004-6361/202245377},
  \href {https://ui.adsabs.harvard.edu/abs/2023A&A...670L..12D} {670, L12}

\bibitem[\protect\citeauthoryear{{Derishev} \& {Piran}}{{Derishev} \&
  {Piran}}{2021}]{2021ApJ923135D}
{Derishev} E.,  {Piran} T.,  2021, \mn@doi [\apj] {10.3847/1538-4357/ac2dec},
  \href {https://ui.adsabs.harvard.edu/abs/2021ApJ...923..135D} {923, 135}

\bibitem[\protect\citeauthoryear{Dichiara, Gropp, Kennea, Kuin, Lien, Marshall,
  Tohuvavohu  \& Williams}{Dichiara et~al.}{2022}]{Dichiara_2022}
Dichiara S.,  Gropp J.~D.,  Kennea J.~A.,  Kuin N. P.~M.,  Lien A.~Y.,
  Marshall F.~E.,  Tohuvavohu A.,   Williams M.~A.,  2022, GCN Circ., 32632

\bibitem[\protect\citeauthoryear{{Fan}, {Piran}, {Narayan}  \& {Wei}}{{Fan}
  et~al.}{2008}]{2008MNRAS1483F}
{Fan} Y.-Z.,  {Piran} T.,  {Narayan} R.,   {Wei} D.-M.,  2008, \mn@doi [\mnras]
  {10.1111/j.1365-2966.2007.12765.x}, \href
  {https://ui.adsabs.harvard.edu/abs/2008MNRAS.384.1483F} {384, 1483}

\bibitem[\protect\citeauthoryear{{Finke} \& {Razzaque}}{{Finke} \&
  {Razzaque}}{2023}]{2023ApJ...942L..21F}
{Finke} J.~D.,  {Razzaque} S.,  2023, \mn@doi [\apjl]
  {10.3847/2041-8213/acade1}, \href
  {https://ui.adsabs.harvard.edu/abs/2023ApJ...942L..21F} {942, L21}

\bibitem[\protect\citeauthoryear{{Finke}, {Razzaque}  \& {Dermer}}{{Finke}
  et~al.}{2010}]{Finke2010}
{Finke} J.~D.,  {Razzaque} S.,   {Dermer} C.~D.,  2010, \mn@doi [\apj]
  {10.1088/0004-637X/712/1/238}, \href
  {https://ui.adsabs.harvard.edu/abs/2010ApJ...712..238F} {712, 238}

\bibitem[\protect\citeauthoryear{{Fraija} et~al.,}{{Fraija}
  et~al.}{2019}]{Fraija2019}
{Fraija} N.,  et~al., 2019, \mn@doi [\apj] {10.3847/1538-4357/ab3e4b}, \href
  {https://ui.adsabs.harvard.edu/abs/2019ApJ...885...29F} {885, 29}

\bibitem[\protect\citeauthoryear{{Fraija}, {Veres}, {Beniamini},
  {Galvan-Gamez}, {Metzger}, {Barniol Duran}  \& {Becerra}}{{Fraija}
  et~al.}{2021}]{2021ApJ91812F}
{Fraija} N.,  {Veres} P.,  {Beniamini} P.,  {Galvan-Gamez} A.,  {Metzger}
  B.~D.,  {Barniol Duran} R.,   {Becerra} R.~L.,  2021, \mn@doi [\apj]
  {10.3847/1538-4357/ac0aed}, \href
  {https://ui.adsabs.harvard.edu/abs/2021ApJ...918...12F} {918, 12}

\bibitem[\protect\citeauthoryear{{Galanti}, {Roncadelli}  \&
  {Tavecchio}}{{Galanti} et~al.}{2022}]{Galanti2022}
{Galanti} G.,  {Roncadelli} M.,   {Tavecchio} F.,  2022, \mn@doi [arXiv
  e-prints] {10.48550/arXiv.2211.06935}, \href
  {https://ui.adsabs.harvard.edu/abs/2022arXiv221106935G} {p. arXiv:2211.06935}

\bibitem[\protect\citeauthoryear{{Gehrels} \& {Razzaque}}{{Gehrels} \&
  {Razzaque}}{2013}]{2013FrPhy...8..661G}
{Gehrels} N.,  {Razzaque} S.,  2013, \mn@doi [Frontiers of Physics]
  {10.1007/s11467-013-0282-3}, \href
  {https://ui.adsabs.harvard.edu/abs/2013FrPhy...8..661G} {8, 661}

\bibitem[\protect\citeauthoryear{{Georganopoulos}, {Kirk}  \&
  {Mastichiadis}}{{Georganopoulos} et~al.}{2001}]{GKM2001}
{Georganopoulos} M.,  {Kirk} J.~G.,   {Mastichiadis} A.,  2001, \apj, 561, 111

\bibitem[\protect\citeauthoryear{{Ghisellini} et~al.,}{{Ghisellini}
  et~al.}{2020}]{2020A&A...636A..82G}
{Ghisellini} G.,  et~al., 2020, \mn@doi [\aap] {10.1051/0004-6361/201937244},
  \href {https://ui.adsabs.harvard.edu/abs/2020A&A...636A..82G} {636, A82}

\bibitem[\protect\citeauthoryear{{Gill} \& {Granot}}{{Gill} \&
  {Granot}}{2022}]{2022Galax..10...74G}
{Gill} R.,  {Granot} J.,  2022, \mn@doi [Galaxies] {10.3390/galaxies10030074},
  \href {https://ui.adsabs.harvard.edu/abs/2022Galax..10...74G} {10, 74}

\bibitem[\protect\citeauthoryear{{Huang}, {Kirk}, {Giacinti}  \&
  {Reville}}{{Huang} et~al.}{2022a}]{2022ApJ...925..182H}
{Huang} Z.-Q.,  {Kirk} J.~G.,  {Giacinti} G.,   {Reville} B.,  2022a, \mn@doi
  [\apj] {10.3847/1538-4357/ac3f38}, \href
  {https://ui.adsabs.harvard.edu/abs/2022ApJ...925..182H} {925, 182}

\bibitem[\protect\citeauthoryear{Huang, Hu, Chen, Zha, Liu, Yao  \& Cao}{Huang
  et~al.}{2022b}]{Huang_2022}
Huang Y.,  Hu S.,  Chen S.,  Zha M.,  Liu C.,  Yao Z.,   Cao Z.,  2022b, GCN
  Circ., 32677

\bibitem[\protect\citeauthoryear{{Inoue}, {Inoue}, {Kobayashi}, {Makiya},
  {Niino}  \& {Totani}}{{Inoue} et~al.}{2013}]{Inoue2013}
{Inoue} Y.,  {Inoue} S.,  {Kobayashi} M. A.~R.,  {Makiya} R.,  {Niino} Y.,
  {Totani} T.,  2013, \mn@doi [\apj] {10.1088/0004-637X/768/2/197}, \href
  {https://ui.adsabs.harvard.edu/abs/2013ApJ...768..197I} {768, 197}

\bibitem[\protect\citeauthoryear{{Isravel}, {Pe'er}  \& {Begue}}{{Isravel}
  et~al.}{2022}]{2022arXiv221002363I}
{Isravel} H.,  {Pe'er} A.,   {Begue} D.,  2022, \mn@doi [arXiv e-prints]
  {10.48550/arXiv.2210.02363}, \href
  {https://ui.adsabs.harvard.edu/abs/2022arXiv221002363I} {p. arXiv:2210.02363}

\bibitem[\protect\citeauthoryear{{Joshi} \& {Razzaque}}{{Joshi} \&
  {Razzaque}}{2021}]{Joshi2021}
{Joshi} J.~C.,  {Razzaque} S.,  2021, \mnras, 505, 1718

\bibitem[\protect\citeauthoryear{{Klinger}, {Tak}, {Taylor}  \&
  {Zhu}}{{Klinger} et~al.}{2023}]{2023MNRAS.520..839K}
{Klinger} M.,  {Tak} D.,  {Taylor} A.~M.,   {Zhu} S.~J.,  2023, \mn@doi
  [\mnras] {10.1093/mnras/stad142}, \href
  {https://ui.adsabs.harvard.edu/abs/2023MNRAS.520..839K} {520, 839}

\bibitem[\protect\citeauthoryear{{Kumar} \& {Zhang}}{{Kumar} \&
  {Zhang}}{2015}]{2015PhR...561....1K}
{Kumar} P.,  {Zhang} B.,  2015, \mn@doi [\physrep]
  {10.1016/j.physrep.2014.09.008}, \href
  {https://ui.adsabs.harvard.edu/abs/2015PhR...561....1K} {561, 1}

\bibitem[\protect\citeauthoryear{{LHAASO Collaboration} et~al.,}{{LHAASO
  Collaboration} et~al.}{2023}]{LHAASO2023Sci}
{LHAASO Collaboration} et~al., 2023, \mn@doi [Science]
  {10.1126/science.adg9328}, \href
  {https://ui.adsabs.harvard.edu/abs/2023Sci...380.1390L} {380, 1390}

\bibitem[\protect\citeauthoryear{{Laskar} et~al.,}{{Laskar}
  et~al.}{2023}]{2023ApJ...946L..23L}
{Laskar} T.,  et~al., 2023, \mn@doi [\apjl] {10.3847/2041-8213/acbfad}, \href
  {https://ui.adsabs.harvard.edu/abs/2023ApJ...946L..23L} {946, L23}

\bibitem[\protect\citeauthoryear{{Lesage} et~al.,}{{Lesage}
  et~al.}{2023}]{2023arXiv230314172L}
{Lesage} S.,  et~al., 2023, \mn@doi [arXiv e-prints]
  {10.48550/arXiv.2303.14172}, \href
  {https://ui.adsabs.harvard.edu/abs/2023arXiv230314172L} {p. arXiv:2303.14172}

\bibitem[\protect\citeauthoryear{{Li} \& {Ma}}{{Li} \&
  {Ma}}{2023}]{2023APh...14802831L}
{Li} H.,  {Ma} B.-Q.,  2023, \mn@doi [Astroparticle Physics]
  {10.1016/j.astropartphys.2023.102831}, \href
  {https://ui.adsabs.harvard.edu/abs/2023APh...14802831L} {148, 102831}

\bibitem[\protect\citeauthoryear{{Li} \& {Wang}}{{Li} \&
  {Wang}}{2023}]{2023arXiv230206116L}
{Li} L.,  {Wang} Y.,  2023, \mn@doi [arXiv e-prints]
  {10.48550/arXiv.2302.06116}, \href
  {https://ui.adsabs.harvard.edu/abs/2023arXiv230206116L} {p. arXiv:2302.06116}

\bibitem[\protect\citeauthoryear{{Liu}, {Zhang}  \& {Wang}}{{Liu}
  et~al.}{2023}]{2023ApJ943L2L}
{Liu} R.-Y.,  {Zhang} H.-M.,   {Wang} X.-Y.,  2023, \mn@doi [\apjl]
  {10.3847/2041-8213/acaf5e}, \href
  {https://ui.adsabs.harvard.edu/abs/2023ApJ...943L...2L} {943, L2}

\bibitem[\protect\citeauthoryear{{MAGIC Collaboration} et~al.,}{{MAGIC
  Collaboration} et~al.}{2019a}]{2019Natur575455M}
{MAGIC Collaboration} et~al., 2019a, \mn@doi [\nat]
  {10.1038/s41586-019-1750-x}, \href
  {https://ui.adsabs.harvard.edu/abs/2019Natur.575..455M} {575, 455}

\bibitem[\protect\citeauthoryear{{MAGIC Collaboration} et~al.,}{{MAGIC
  Collaboration} et~al.}{2019b}]{2019Natur575459M}
{MAGIC Collaboration} et~al., 2019b, \mn@doi [\nat]
  {10.1038/s41586-019-1754-6}, \href
  {https://ui.adsabs.harvard.edu/abs/2019Natur.575..459M} {575, 459}

\bibitem[\protect\citeauthoryear{{Mao} \& {Wang}}{{Mao} \&
  {Wang}}{2023}]{2023ApJ946.89M}
{Mao} J.,  {Wang} J.,  2023, \mn@doi [\apj] {10.3847/1538-4357/acc400}, \href
  {https://ui.adsabs.harvard.edu/abs/2023ApJ...946...89M} {946, 89}

\bibitem[\protect\citeauthoryear{{M{\'e}sz{\'a}ros}}{{M{\'e}sz{\'a}ros}}{2006}]{2006RPPh...69.2259M}
{M{\'e}sz{\'a}ros} P.,  2006, \mn@doi [Reports on Progress in Physics]
  {10.1088/0034-4885/69/8/R01}, \href
  {https://ui.adsabs.harvard.edu/abs/2006RPPh...69.2259M} {69, 2259}

\bibitem[\protect\citeauthoryear{{Nakagawa}, {Takahashi}, {Yamada}  \&
  {Yin}}{{Nakagawa} et~al.}{2023}]{2023PhLB..83937824N}
{Nakagawa} S.,  {Takahashi} F.,  {Yamada} M.,   {Yin} W.,  2023, \mn@doi
  [Physics Letters B] {10.1016/j.physletb.2023.137824}, \href
  {https://ui.adsabs.harvard.edu/abs/2023PhLB..83937824N} {839, 137824}

\bibitem[\protect\citeauthoryear{{Nakar}, {Ando}  \& {Sari}}{{Nakar}
  et~al.}{2009}]{2009ApJ_Nakar_KN}
{Nakar} E.,  {Ando} S.,   {Sari} R.,  2009, \mn@doi [\apj]
  {10.1088/0004-637X/703/1/675}, \href
  {https://ui.adsabs.harvard.edu/abs/2009ApJ...703..675N} {703, 675}

\bibitem[\protect\citeauthoryear{{Piran}}{{Piran}}{2004}]{2004RvMP...76.1143P}
{Piran} T.,  2004, \mn@doi [Reviews of Modern Physics]
  {10.1103/RevModPhys.76.1143}, \href
  {https://ui.adsabs.harvard.edu/abs/2004RvMP...76.1143P} {76, 1143}

\bibitem[\protect\citeauthoryear{{Razzaque}}{{Razzaque}}{2013}]{2013PhRvD..88j3003R}
{Razzaque} S.,  2013, \mn@doi [\prd] {10.1103/PhysRevD.88.103003}, \href
  {https://ui.adsabs.harvard.edu/abs/2013PhRvD..88j3003R} {88, 103003}

\bibitem[\protect\citeauthoryear{{Razzaque}, {Dermer}  \& {Finke}}{{Razzaque}
  et~al.}{2010}]{2010OAJ.....3..150R}
{Razzaque} S.,  {Dermer} C.~D.,   {Finke} J.~D.,  2010, \mn@doi [The Open
  Astronomy Journal] {10.2174/1874381101003010150}, \href
  {https://ui.adsabs.harvard.edu/abs/2010OAJ.....3..150R} {3, 150}

\bibitem[\protect\citeauthoryear{{Ren}, {Wang}, {Zhang}  \& {Dai}}{{Ren}
  et~al.}{2023}]{2023ApJ...947...53R}
{Ren} J.,  {Wang} Y.,  {Zhang} L.-L.,   {Dai} Z.-G.,  2023, \mn@doi [\apj]
  {10.3847/1538-4357/acc57d}, \href
  {https://ui.adsabs.harvard.edu/abs/2023ApJ...947...53R} {947, 53}

\bibitem[\protect\citeauthoryear{{Ronchi} et~al.,}{{Ronchi}
  et~al.}{2020}]{2020A&A...636A..55R}
{Ronchi} M.,  et~al., 2020, \mn@doi [\aap] {10.1051/0004-6361/201936765}, \href
  {https://ui.adsabs.harvard.edu/abs/2020A&A...636A..55R} {636, A55}

\bibitem[\protect\citeauthoryear{{Sahu} \& {Fort{\'\i}n}}{{Sahu} \&
  {Fort{\'\i}n}}{2020}]{2020ApJ...895L..41S}
{Sahu} S.,  {Fort{\'\i}n} C. E.~L.,  2020, \mn@doi [\apjl]
  {10.3847/2041-8213/ab93da}, \href
  {https://ui.adsabs.harvard.edu/abs/2020ApJ...895L..41S} {895, L41}

\bibitem[\protect\citeauthoryear{{Sahu}, {Valadez Polanco}  \&
  {Rajpoot}}{{Sahu} et~al.}{2022}]{2022ApJ92970S}
{Sahu} S.,  {Valadez Polanco} I.~A.,   {Rajpoot} S.,  2022, \mn@doi [\apj]
  {10.3847/1538-4357/ac5cc6}, \href
  {https://ui.adsabs.harvard.edu/abs/2022ApJ...929...70S} {929, 70}

\bibitem[\protect\citeauthoryear{{Salafia} et~al.,}{{Salafia}
  et~al.}{2022}]{Salafia2022}
{Salafia} O.~S.,  et~al., 2022, \mn@doi [\apjl] {10.3847/2041-8213/ac6c28},
  \href {https://ui.adsabs.harvard.edu/abs/2022ApJ...931L..19S} {931, L19}

\bibitem[\protect\citeauthoryear{{Sari} \& {Esin}}{{Sari} \&
  {Esin}}{2001}]{2001ApJ548787S}
{Sari} R.,  {Esin} A.~A.,  2001, \mn@doi [\apj] {10.1086/319003}, \href
  {https://ui.adsabs.harvard.edu/abs/2001ApJ...548..787S} {548, 787}

\bibitem[\protect\citeauthoryear{{Sari}, {Piran}  \& {Narayan}}{{Sari}
  et~al.}{1998}]{1998ApJ...497L..17S}
{Sari} R.,  {Piran} T.,   {Narayan} R.,  1998, \mn@doi [\apjl]
  {10.1086/311269}, \href
  {https://ui.adsabs.harvard.edu/abs/1998ApJ...497L..17S} {497, L17}

\bibitem[\protect\citeauthoryear{{Sato}, {Obayashi}, {Yamazaki}, {Murase}  \&
  {Ohira}}{{Sato} et~al.}{2021}]{2021MNRAS.504.5647S}
{Sato} Y.,  {Obayashi} K.,  {Yamazaki} R.,  {Murase} K.,   {Ohira} Y.,  2021,
  \mn@doi [\mnras] {10.1093/mnras/stab1273}, \href
  {https://ui.adsabs.harvard.edu/abs/2021MNRAS.504.5647S} {504, 5647}

\bibitem[\protect\citeauthoryear{{Sato}, {Murase}, {Ohira}  \&
  {Yamazaki}}{{Sato} et~al.}{2023}]{2023MNRAS.522L..56S}
{Sato} Y.,  {Murase} K.,  {Ohira} Y.,   {Yamazaki} R.,  2023, \mn@doi [\mnras]
  {10.1093/mnrasl/slad038}, \href
  {https://ui.adsabs.harvard.edu/abs/2023MNRAS.522L..56S} {522, L56}

\bibitem[\protect\citeauthoryear{{Smirnov} \& {Trautner}}{{Smirnov} \&
  {Trautner}}{2022}]{Smirnov2022arXiv221100634S}
{Smirnov} A.~Y.,  {Trautner} A.,  2022, arXiv e-prints, \href
  {https://ui.adsabs.harvard.edu/abs/2022arXiv221100634S} {p. arXiv:2211.00634}

\bibitem[\protect\citeauthoryear{Veres, Burns, Bissaldi, Lesage  \&
  Roberts}{Veres et~al.}{2022}]{Veres_2022}
Veres P.,  Burns E.,  Bissaldi E.,  Lesage S.,   Roberts O.,  2022, GCN Circ.,
  32636

\bibitem[\protect\citeauthoryear{{Wang} et~al.,}{{Wang}
  et~al.}{2015}]{2015ApJS..219....9W}
{Wang} X.-G.,  et~al., 2015, \mn@doi [\apjs] {10.1088/0067-0049/219/1/9}, \href
  {https://ui.adsabs.harvard.edu/abs/2015ApJS..219....9W} {219, 9}

\bibitem[\protect\citeauthoryear{{Wang}, {Liu}, {Zhang}, {Xi}  \&
  {Zhang}}{{Wang} et~al.}{2019}]{2019ApJ...884..117W}
{Wang} X.-Y.,  {Liu} R.-Y.,  {Zhang} H.-M.,  {Xi} S.-Q.,   {Zhang} B.,  2019,
  \mn@doi [\apj] {10.3847/1538-4357/ab426c}, \href
  {https://ui.adsabs.harvard.edu/abs/2019ApJ...884..117W} {884, 117}

\bibitem[\protect\citeauthoryear{{Wang}, {Xi}, {Liu}, {Xue}  \& {Wang}}{{Wang}
  et~al.}{2020}]{2020PhRvD.101h3004W}
{Wang} Z.-R.,  {Xi} S.-Q.,  {Liu} R.-Y.,  {Xue} R.,   {Wang} X.-Y.,  2020,
  \mn@doi [\prd] {10.1103/PhysRevD.101.083004}, \href
  {https://ui.adsabs.harvard.edu/abs/2020PhRvD.101h3004W} {101, 083004}

\bibitem[\protect\citeauthoryear{{Williams} et~al.,}{{Williams}
  et~al.}{2023}]{2023ApJ24W}
{Williams} M.~A.,  et~al., 2023, \mn@doi [\apjl] {10.3847/2041-8213/acbcd1},
  \href {https://ui.adsabs.harvard.edu/abs/2023ApJ...946L..24W} {946, L24}

\bibitem[\protect\citeauthoryear{{Yamasaki} \& {Piran}}{{Yamasaki} \&
  {Piran}}{2022}]{yama2022MNRAS142Y}
{Yamasaki} S.,  {Piran} T.,  2022, \mn@doi [\mnras] {10.1093/mnras/stac483},
  \href {https://ui.adsabs.harvard.edu/abs/2022MNRAS.512.2142Y} {512, 2142}

\bibitem[\protect\citeauthoryear{{Zhang} \& {M{\'e}sz{\'a}ros}}{{Zhang} \&
  {M{\'e}sz{\'a}ros}}{2001}]{2001ApJ...559..110Z}
{Zhang} B.,  {M{\'e}sz{\'a}ros} P.,  2001, \mn@doi [\apj] {10.1086/322400},
  \href {https://ui.adsabs.harvard.edu/abs/2001ApJ...559..110Z} {559, 110}

\bibitem[\protect\citeauthoryear{{Zhang}, {Ren}, {Huang}, {Liang}, {Lin}  \&
  {Liang}}{{Zhang} et~al.}{2021a}]{2021ApJ...917...95Z}
{Zhang} L.-L.,  {Ren} J.,  {Huang} X.-L.,  {Liang} Y.-F.,  {Lin} D.-B.,
  {Liang} E.-W.,  2021a, \mn@doi [\apj] {10.3847/1538-4357/ac0c7f}, \href
  {https://ui.adsabs.harvard.edu/abs/2021ApJ...917...95Z} {917, 95}

\bibitem[\protect\citeauthoryear{{Zhang}, {Murase}, {Veres}  \&
  {M{\'e}sz{\'a}ros}}{{Zhang} et~al.}{2021b}]{2021ApJ92055Z}
{Zhang} B.~T.,  {Murase} K.,  {Veres} P.,   {M{\'e}sz{\'a}ros} P.,  2021b,
  \mn@doi [\apj] {10.3847/1538-4357/ac0cfc}, \href
  {https://ui.adsabs.harvard.edu/abs/2021ApJ...920...55Z} {920, 55}

\bibitem[\protect\citeauthoryear{{Zhang}, {Murase}, {Ioka}, {Song}, {Yuan}  \&
  {M{\'e}sz{\'a}ros}}{{Zhang} et~al.}{2023}]{2023ApJ947L14Z}
{Zhang} B.~T.,  {Murase} K.,  {Ioka} K.,  {Song} D.,  {Yuan} C.,
  {M{\'e}sz{\'a}ros} P.,  2023, \mn@doi [\apjl] {10.3847/2041-8213/acc79f},
  \href {https://ui.adsabs.harvard.edu/abs/2023ApJ...947L..14Z} {947, L14}

\makeatother
\end{thebibliography}



\bsp	
\label{lastpage}
\end{document}